\documentclass[runningheads,epj]{svjour}

\usepackage{widetext}

\usepackage{amssymb}
\usepackage{amsfonts}
\usepackage{bm}
\usepackage{epsfig}
\usepackage{epstopdf}
\usepackage{epsf}
\usepackage[]{graphicx}
\usepackage{wrapfig}
\usepackage{color}
\usepackage{cite}
\usepackage{xspace}

\begin{document}


\title
{Ab-initio modeling of an anion $C_{60}^-$ pseudopotential  for fullerene-based compounds.}

\author{
I I Vrubel\inst{1},
R G Polozkov\inst{1,2},
and 
V K Ivanov\inst{3}}

\institute{ITMO University, 49 Kronverksky Pr., St. Petersburg, 197101, Russia \and 
Science Institute, University of Iceland, Dunhagi 3,
IS-107, Reykjavik, Iceland \and  Peter the Great Saint-Petersburg Polytechnic University,  29 Politekhnicheskaya, St. Petersburg, 195251, Russia}
%

%
\date{Received: date / Revised version: date}
%
\abstract{
A pseudopotential of $C_{60}^-$ has been constructed from $\emph{ab-initio}$ quantum-mechanical calculations. 
Since the obtained pseudopotential can be easily fitted by rather simple analytical approximation it can be effectively used both 
in classical and quantum molecular dynamics of fullerene-based compounds.
}


\authorrunning{R G Polozkov et al}
\titlerunning{Pseudopotential of an anion $C_{60}^-$}
\maketitle
\section{Introduction}
\textcolor{black}{
Fullerene $C_{60}$ is the most studied and widely used among all fullerenes because of the availability, high symmetry and low price \cite{Chem_of_full}. 
Some of the most promising fields of application of these novel materials are artificial photosynthesis, non-linear optics and the preparation of photoactive films 
and nanostructures (see, for example \cite{Fullerenes_Application}). 
Because of high electron affinity and small rearrangement energy fullerenes, in particular $C_{60}$, 
play a role of electron-acceptors in such systems and produce very stable radical pairs.  In particular, it was demonstrated that a $\pi$-conjugated polymer 
was able to efficiently transfer electrons to the $C_{60}$ core giving rise to long-lived charge-separated states. 
For example the donor-acceptor compound [$C_{60}$]PCBM \cite{PCBM} is the most known and effectively used in organic solar photoelectric cells 
for the last time.
But, due to the difficulties in the modeling extended and 
possible nanostructured materials as [$C_{60}$]PCBM an isolated molecular limit is preferable, so the investigations of the isolated anions and radical anions of fullerenes, 
in particular $C_{60}^-$, seems to be actual.  
}

\textcolor{black}{
The anions and radical anions of fullerenes have been an object of 
intensively investigations during last two decade \cite{Chem_of_full, Wang_book, Cataldo}. The anionic fullerenes
have been observed in ion cyclotron resonance traps \cite{Hettich}, storage rings \cite{Tomita} and electrospray mass spectrometry \cite{Wang_5}. 
The theoretical systematic study of the stability of highly charged anionic fullerenes 
has been performed  
within different levels of theory \cite{Green, Wang}. But for the developing efficient quantum simulation methods, 
which allow us to predict the optimized geometry of the fullerene $C_{60}$ compounds with reasonable computer cost and accuracy, we suggest to construct a pseudopotential of $C_{60}^-$. 
}

\textcolor{black}{
The recent calculations showed that the application of simple and widely 
used jellium model doesn't bring data into accordance with results of more complicated but 
accurate \emph{ab-initio} calculations  \cite{Verkhovtsev}. In this paper the pseudopotential of $C_{60}^-$ has been constructed on the basis of the total electrostatic potential 
of  $C_{60}^-$ calculated 
within the \emph{ab-initio} approach.
}

\textcolor{black}{
The atomic system of units, $m = |e| = \hbar = 1$, is used throughout the paper.
}
\section{Method of calculation \label{Method}}

\subsection{\emph{Ab-initio} calculations
\label{Ab-initio calculations}}

\textcolor{black}{
All \emph{ab-initio} computations 
are performed 
by using the FireFly QC package \cite{Firefly}. For the first the fully optimized geometry and the total energy of $C_{60}^-$ 
have been obtained 
from 
the 
Hartree-Fock and density functional theory (DFT) calculations by ROHF/6-31G(d) and B3LYP/6-31G(d) levels respectively. 
Then within the optimized geometry the \emph{ab-initio} calculations of the electronic structure 
and the total charge density of $C_{60}^-$ 
have been performed 
at the same levels of theory, which 
is shown 
to provide reasonable results for small carbon clusters \cite{Alcami1} and 
fullerenes \cite{Wang, Alcami2}, both charged and neutral.  
Although the inclusion of diffuse functions is usually important to obtain accurate absolute energies for anions,
it has recently been shown that the 6-31G(d) and 6-31G+(d) basis sets give similar results for geometries,
charge distributions, and relative energies of anionic $C_{60}$ and $C_{70}$ fullerenes \cite{Zetter1, Zetter2}.
}

\textcolor{black}{
The important point of the \emph{ab-initio} calculations of the total charge density 
is to apply  
the corresponding key in the input of the FireFly program like $AIMPAC=1$ to obtain 
the practical information about molecular orbital wave functions which 
are used 
in the next step of construction of the pseudopotential of $C_{60}^-$.
}

\subsection{Pseudopotential construction
\label{Pseudopotential}}
\textcolor{black}{
The  
pseudopotential of $C_{60}^-$ can be construct on the basis of the total electrostatic potential. 
The latter 
is presented as a sum of two summands: 
the potential of nuclei $U_{\rm n}(\textbf r)$, which depends on positions of sixty carbon atoms, and the potential created by electron density $\rho(\textbf{r})$ of 361 electrons $U_{\rm el}(\textbf r)$: 
}

\begin{eqnarray}
\label{pot1}
&U_{\rm tot}&(\textbf r)=U_{\rm n}(\textbf r)+U_{\rm el}(\textbf r) = \nonumber \\
&-&\sum_{\rm i=1}^{60}\frac{6}{|\textbf r - \textbf R_{\rm i}|}+
\int\frac{\rho(\textbf{r}')}{|\textbf{r} - \textbf{r}'|}d\textbf{r}' \ .
\end{eqnarray}

\begin{figure}[h]
\flushright
\includegraphics[scale=0.42,clip]{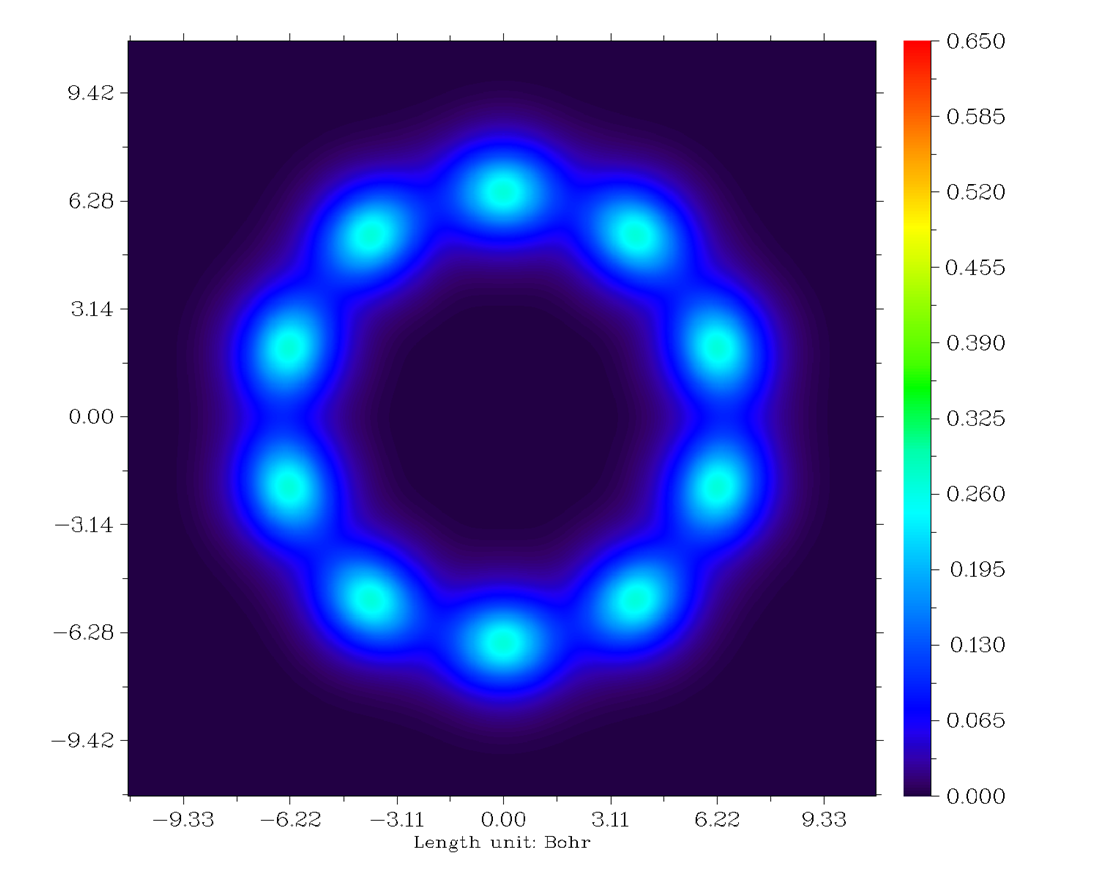}  
\vspace{5ex}
\caption{
A color filled map of the electron charge density of the $C_{60}^-$ obtained from \emph{ab-initio} calculations prepared within the Multiwfn \cite{Multiwfn} software in the plane X,Y. 
The origin of the X,Y,Z is set at the center of $C_{60}^-$, lengths unit is Bohr.  
}
\label{Figure.01} 
\end{figure}

\textcolor{black}{
The positions of the carbon atoms within the optimized geometry and corresponding charge density have been extracted from results 
of the \emph{ab-initio} FireFly QC package calculations by using of 
a Multifunctional Wavefunction Analyzer (Multiwfn) \cite{Multiwfn} (see for example a color filled map of the electron charge density 
of the $C_{60}^-$ obtained from \emph{ab-initio} calculations prepared within the Multiwfn \cite{Multiwfn} software on the Fig. \ref{Figure.01}). 
This software has been then used for 
computation of the corresponding electrostatic potentials on a specified grid of the position vector $\textbf r$.
After that we have averaged the electrostatic potential obtained from Multiwfn software over the directions 
of the position vector $\textbf r$ to construct the radial dependence of $C_{60}^-$ pseudopotential $U_{\rm pseudo}(r)$ and 
to obtain 
averaged electron density $\overline{\rho}(r)$:
}

\begin{eqnarray}
&U_{\rm pseudo}(r)&=\overline{U}_{\rm tot}(r) = \overline{U}_{\rm n}(r) + \overline{U}_{\rm el}(r) \ , \nonumber \\
&\overline{U}_{\rm i}(r)& = \frac{1}{4\pi} \int U_{\rm i}(\textbf r) d\Omega \quad (\rm i = tot, n, el) \ , \nonumber \\
&\overline{\rho}(r)& = \frac{1}{4\pi} \int \rho(\textbf r) d\Omega \ .
\end{eqnarray}

\section{Results 
\label{Results}}
\begin{figure} [h]
\includegraphics[scale=0.35,clip]{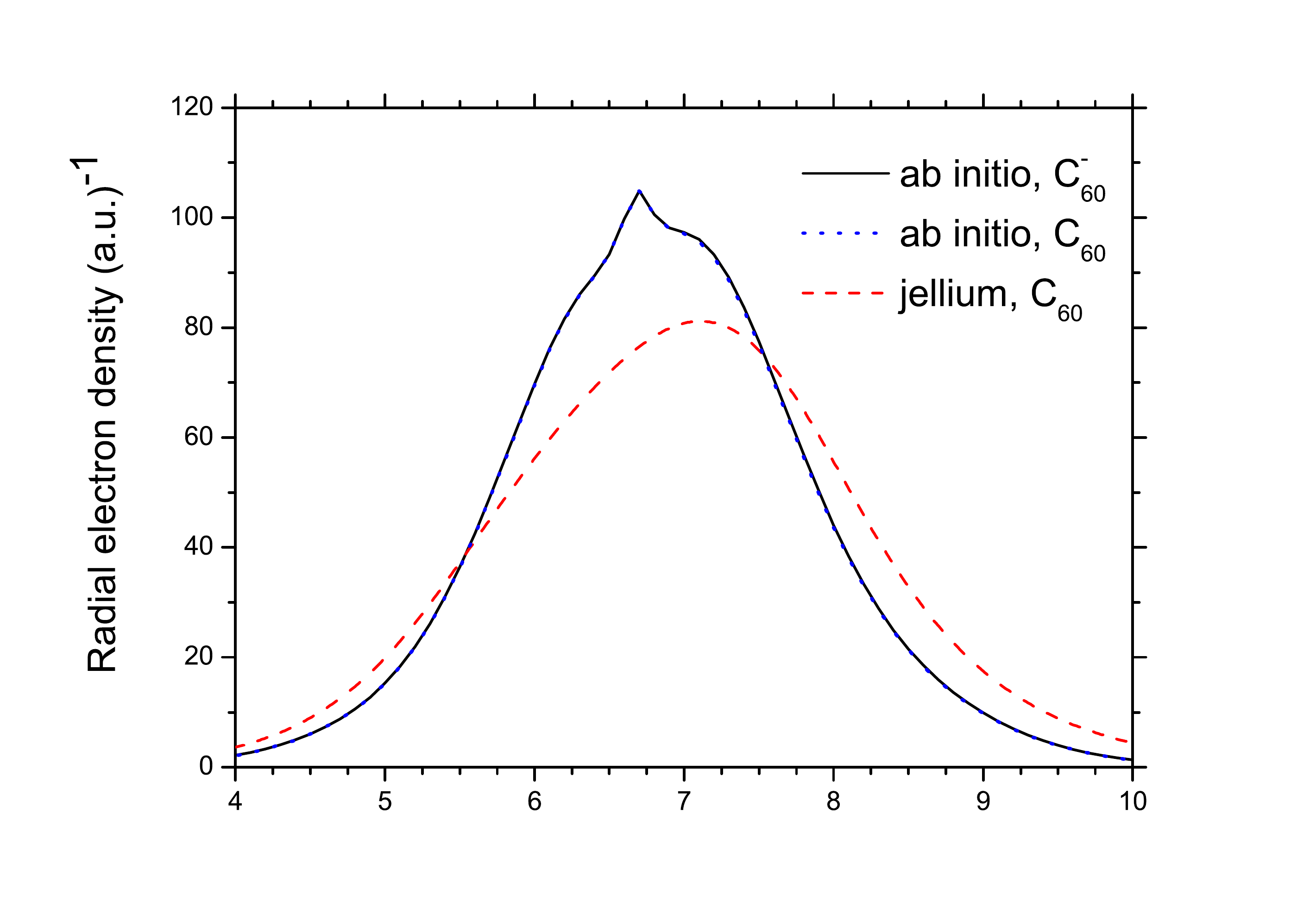}  
\caption{
Radial valence electron density of $C_{60}^-$(ROHF/6-31G(d) level, solid line) is compared with the same of $C_{60}$ calculated by \emph{ab-initio} method \cite{Verkhovtsev}(ROHF/6-31G(d) level, blue dotted line) and with use 
the jellium model \cite{Verkhovtsev} (red dashed line). 
}
\label{Figure.02} 
\end{figure}

\textcolor{black}{
For the first we have checked the non-applicability of the jellium model for purpose of construction of pseudopotential of $C_{60}^-$.
The averaged radial valence electron density $\overline{\rho}(r)$ of $C_{60}^-$ calculated by \emph{ab-initio} method 
has been compared with  
results of the jellium model and \emph{ab-initio} calculations for $C_{60}$\cite{Verkhovtsev}.  
The Hartree-Fock method 
has been used 
for self-consistent calculations 
in all three cases. Fig.\ref{Figure.02} demonstrates the density profiles of valence electrons of $C_{60}$ \cite{Verkhovtsev} 
and $C_{60}^-$ (present work, ROHF/6-31G(d) level) as a function of radial distance from a center of fullerene. 
As Fig.\ref{Figure.02} indicates, the results of \emph{ab-initio} calculations for fullerene and anion are close, but substantially differ
from the results of jellium model calculations for $C_{60}$, which makes this approach non applicable to solving a problem 
of determination of $C_{60}^-$ pseudopotential.
}

\begin{figure} [h]
\centering
\includegraphics[scale=0.35,clip]{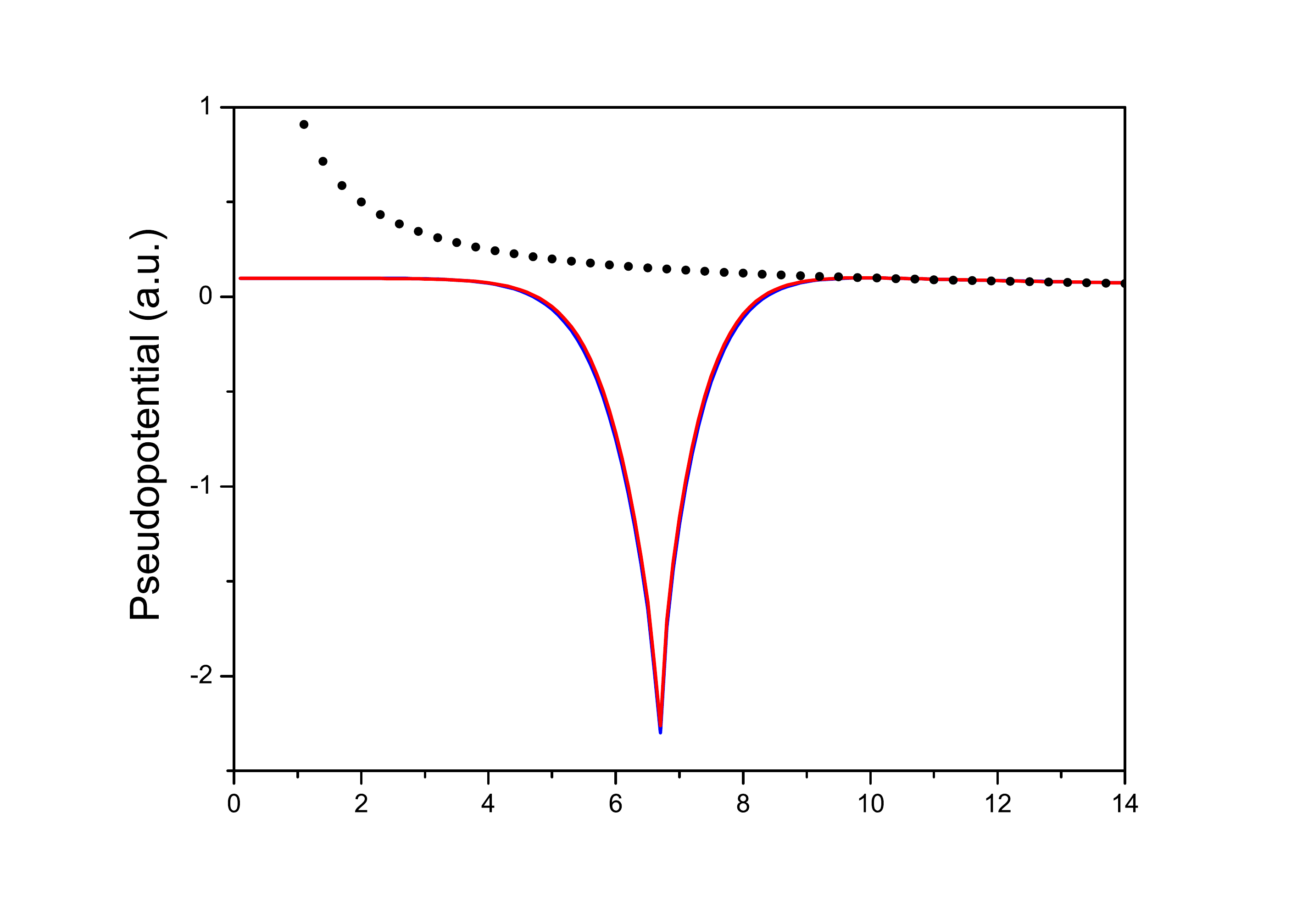}  
\caption{
Pseudopotential of $C_{60}^-$  obtained from \emph{ab-initio} calculations: ROHF/6-31G(d) (blue solid line), B3LYP/6-31G(d) (red solid line) and compared with an 1/r asymptotic behavior (black dots). 
}
\label{Figure.03} 
\end{figure}

\textcolor{black}{
The results of the $C_{60}^-$ pseudopotential calculations by Hartree-Fock 
method (ROHF/6-31G(d)) and within DFT (B3LYP/6-31G(d)) are presented and compared in Fig.\ref{Figure.03}. 
Note that the usage of 
the different approaches for the electronic structure calculations leads to the significant discrepancy of the corresponding one-particle energies results but 
doesn't lead to the any noticeable differences in the resulting 
behavior of 
pseudopotential (compare blue and red solid line in Fig.3 ).  
}

\textcolor{black}{
It should be 
mentioned 
several important features of the pseudopotential obtained. The first 
one is 
the correct asymptotic behavior at the large distances as $1/r$, which is typical for a single negative ion (see Fig.\ref{Figure.03}). The numerical analysis shows that
the pseudopotential behavior begins to satisfy the $1/r$ law at the radial distance about $10$ a.u. 
Secondly 
the pseudopotential demonstrates two different types of interaction between the fullerene's anion and an external electron: 
the strong attraction close to a radius of fullerene's anion and the weak repulsion outside and inside of fullerene cage. This  
combination of repulsion and attraction gives rise the weak barriers for an any negative projectile particle and can lead to 
increasing of probability for the projectile to "getting stuck" on the fullerene cage. 
}
\vspace{-7ex}

\begin{figure} [h]
\centering
\includegraphics[scale=0.52,clip]{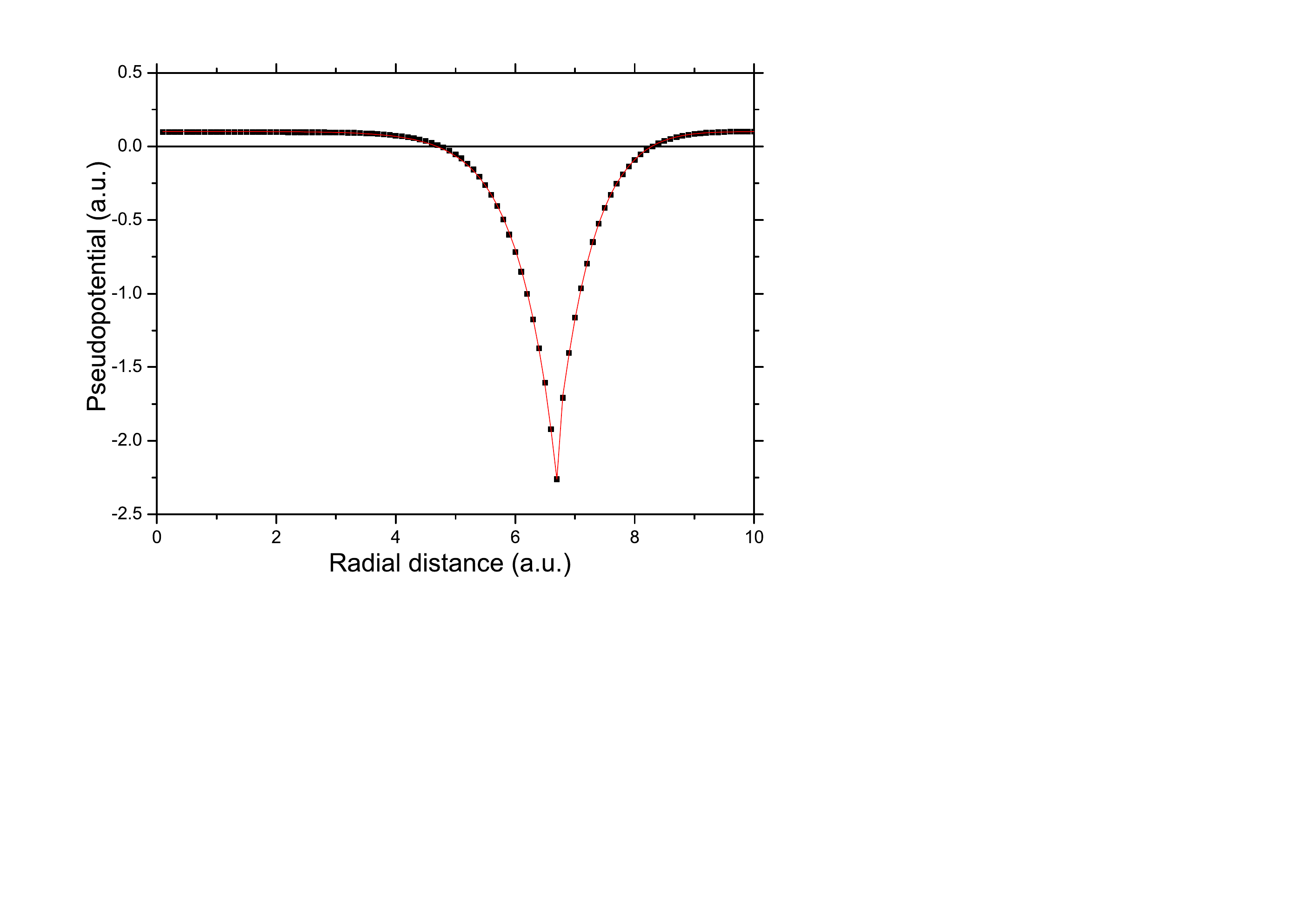}  
\vspace{-30ex}
\caption{
Approximation (red solid line) of the pseudopotential of $C_{60}^-$  obtained from \emph{ab-initio} B3LYP/6-31G(d) calculations (black dots). 
}
\label{Figure.04} 
\end{figure}

\textcolor{black}{
For purposes of the classical and quantum molecular dynamics of fullerene-based compounds it is 
reasonable 
to make the analytical approximation of the numerically obtained pseudopotential. 
Within the range $0-10$ a.u. of radial distance the pseudopotential has been approximated by sum (\ref{cheslercramgenview}) of constant and Chesler-Cram single peak function (see Fig.\ref{Figure.04}).
This Chesler-Cram function \cite{CheslerCram} is 
applied 
to approximate experimental results in the processing of chromatographic data.
We 
use 
this function because it may 
describe 
discontinuity point and consists of elementary functions.
The general view of our function 
is presented by the following formula:
}
\begin{eqnarray}\label{cheslercramgenview}
u(r)&=&y_0+A[
e^{-\frac{(r-r_{c1})^2}{2w}} 
+ Be^{-\frac{1}{2}k_3(|r-r_{c3}|+(r-r_{c3}))}\times \nonumber  \\ 
&&(1-0.5\left(1-\tanh(k_2(r-r_{c2}))\right)~)
],
\end{eqnarray}
\textcolor{black}{
where $r$ is the radial distance, $y_0$, $A$, $r_{c1}$, $w$, $B$, $k_2$, $r_{c2}$, $k_3$, $r_{c3}$ are approximation constants.
The 
array of constants that allows to achieve the best result is represented in 
the 
table \ref{approxconst}.
}

\begin{table}[h]
\centering
\caption{array of constants}
\label{approxconst}
\begin{tabular*}{0.3\textwidth}
{@{\extracolsep{\fill}}ccc}
\hline\noalign{\smallskip}
   $symbol$  & $value$  \\
\noalign{\smallskip}\hline\noalign{\smallskip}
$y_0$      &  0,10096  \\ 
$r_{c1}$  & 6,66092  \\ 
$A$          & -1,60691 \\
$w$         &  $4,0523*10^{-4}$  \\ 
$k_2$      &  0,79686  \\
$r_{c2}$  &  8,32747  \\
$B$          &  20,83852 \\
$k_3$      &  3,14767  \\ 
$r_{c3}$  & 6,66996  \\
\noalign{\smallskip}\hline

\end{tabular*}
\end{table}

\section{\textcolor{black}{Conclusion}}
\textcolor{black}{
In this work we have constructed the pseudopotential of the fullerene anion $C_{60}^-$ for molecular dynamic purposes. 
The method of construction is based on the using of the charge density obtained by the \emph{ab-initio} calculations and 
on the averaging of the corresponding total electrostatic potential to make the radial dependence of the pseudopotential.}

\textcolor{black}{
The pseudopotential of the fullerene anion $C_{60}^-$ obtained has rather simple analytical 
approximation and 
then 
can be effectively used both in classical and quantum molecular dynamics of fullerene-based compounds.
}
\section{Acknowledgments}
\textcolor{black}{
RGP and IIV acknowledge support from Russia's Federal Program "Scientific and Educational Manpower for Innovative Russia" (grant no RFMEFI58715X0020),  RGP acknowledges support by Rannis Project "BOFEHYSS". 
}



\begin{thebibliography}{99}

\bibitem{Chem_of_full}
A. Hirsch, \emph{The Chemistry of the Fullerenes} (WILEY-VCH, 2002) 2.

\bibitem{Fullerenes_Application}
S. Campidelli, A. Mateo-Alonso, and M. Prato, in \emph{Fullerenes: Principles and Applications}, 
edited by F. Langa and J.-F. Nierengarten (RSC Publishing, 2007) chapter 7.

\bibitem{PCBM}
C. J. Brabec, A. Cravino, D. Meissner, N. S. Sariciftci, T. Fromherz, M. T. Rispens, L. Sanchez, and J. C. Hummelen, 
Adv. Funct. Mater. \textbf{11}, (2001) 374.

\bibitem{Cataldo}
F. Cataldo, S. Iglesias-Groth and A. Manchado, Fullerenes, Nanotubes and Carbon Nanostruct. \textbf{21}, (2013) 537.

\bibitem{Wang_book}
Y. Wang, M. Alcami, and F. Martin, in \emph{Handbook of Nanophysics}, edited by K. D. Sattler (Taylor \& Francis, London, 2010) vol.2.

\bibitem{Hettich}
R. L. Hettich, R. N. Compton, and R. H. Ritchie, Phys. Rev. Lett. \textbf{67}, (1991) 1242.

\bibitem{Tomita}
S. Tomita \emph{et al}, J. Chem. Phys. \textbf{124}, (2006) 024310.

\bibitem{Wang_5}
X.-B. Wang, H.-K. Woo, X. Huang, M. M. Kappes, and L. -S. Wang, Phys. Rev. Lett. \textbf{96}, (2006) 143002

\bibitem{Green}
W. H. Green, Jr., S. M. Gorun, G. Fitzgerald, P. W. Fowler, A. Ceulemans and B. C. Titeca, J. Phys. Chem. \textbf{100}, (1996) 14892.  

\bibitem{Wang}
Y. Wang, H. Zettergren, M. Alcami, and F. Martin, Phys. Rev. A \textbf{80}, (2009) 033201. 

\bibitem{Verkhovtsev} A.V. Verkhovtsev, R.G. Polozkov,  V.K. Ivanov, A. V. Korol, A.V. Solov'yov, J. Phys. B: At. Mol. Opt. Phys. \textbf{45}, (2012) 215101.

\bibitem{Firefly} Alex A. Granovsky, Firefly version 8, www http://classic.chem.msu.su/gran/firefly/index.html.

\bibitem{Alcami1}
S. Diaz-Tendero, F. Martin,  and M. Alcami, J. Phys. Chem. A \textbf{106}, (2002) 10782.

\bibitem{Alcami2}
S. Diaz-Tendero, F. Martin,  and M. Alcami, J. Chem. Phys. \textbf{119}, (2003) 5545. 

\bibitem{Zetter1}
H. Zettergren, M. Alcami, and F. Martin, Phys. Rev. A \textbf{76}, (2007) 043205.

\bibitem{Zetter2}
H. Zettergren, M. Alcami, and F. Martin, ChemPhysChem. \textbf{9}, (2008) 861.

\bibitem{Multiwfn} Tian Lu and Feiwu Chen, J. Comput. Chem. \textbf{33}, (2012) 580.

\bibitem{CheslerCram} S.N. Chesler, S.P. Cram, Anal. Chem. \textbf{45(8)}, (1973) 1354. 




  
\end{thebibliography}
\end{document}